\documentclass[aps,showpacs,superscriptaddress]{revtex4}
\usepackage{epsfig}
\usepackage{color}
\newcommand{\ben}{\begin{eqnarray}}
\newcommand{\een}{\end{eqnarray}}
\newcommand{\nnu}{\nonumber\\}
\newcommand{\bef}{\begin{figure}[htb]\centering}
\newcommand{\eef}{\end{figure}}
\newcommand{\bet}{\begin{table}[hbt]\centering}
\newcommand{\eet}{\end{table}}

\begin{document}
\title{Accessing tri-gluon correlations in the nucleon via the \\
single spin asymmetry in open charm production}
\author{Zhong-Bo Kang}
\email{kangzb@iastate.edu}
\affiliation{Department of Physics and Astronomy, 
                 Iowa State University,  
                 Ames, IA 50011, USA}
\author{Jian-Wei Qiu}
\email{jwq@iastate.edu}
\affiliation{Department of Physics and Astronomy, 
                 Iowa State University,  
                 Ames, IA 50011, USA}
\author{Werner Vogelsang} 
\email{vogelsan@quark.phy.bnl.gov}
\affiliation{Physics Department, 
		Brookhaven National Laboratory,
		Upton, NY 11973, USA }
\author{Feng Yuan}
\email{fyuan@lbl.gov}
\affiliation{Nuclear Science Division, 
		Lawrence Berkeley National Laboratory,
		Berkeley, CA 94720, USA}
\affiliation{RIKEN BNL Research Center, Building 510A, 
		Brookhaven National Laboratory, Upton, NY 11973, USA}                                        
\begin{abstract}
We calculate the single transverse-spin asymmetry for open charm production 
in $pp$ collisions within the QCD collinear factorization approach.  
We include contributions from both twist-three quark-gluon and tri-gluon 
correlation functions. We find that the quark-gluon correlation functions 
alone generate only a very small asymmetry for open charm production in 
the kinematic region of current interest at RHIC, so that the observation 
of any significant single-spin asymmetry would be a clear indication of 
the presence of tri-gluon correlations inside a polarized proton. 
We furthermore demonstrate that the tri-gluon contribution could be very 
different for the production of $D$ and $\bar{D}$ mesons. These features 
make the single spin asymmetry in open charm production in polarized 
$pp$ collisions at RHIC an excellent 
probe of tri-gluon correlation functions. 
\end{abstract}                 
\date{\today}
\maketitle

\section{Introduction}
Large single transverse-spin asymmetries (SSAs) 
have been consistently observed in various experiments at different collision energies \cite{SSA-fixed-tgt,SSA-dis,SSA-rhic}. These experimental data have triggered much theoretical activity aiming at identifying
the fundamental mechanism behind the measured asymmetries as well as the relevant components of hadron structure. Among the proposed theoretical mechanisms, the transverse momentum dependent (TMD) parton distribution 
approach \cite{Siv90,MulTanBoe,JiMaYu04,mulders,mulders1,UnifySSA,Qiu:2007ar} and the twist-3 quark-gluon correlation 
approach \cite{Efremov,qiu,koike,koiketanaka} have been
the ones most widely discussed in the literature in recent years.  The 
two approaches each have their kinematic domain of validity.
For semi-inclusive deep inelastic scattering (SIDIS) 
and for the Drell-Yan process, 
they were shown to be consistent with each other in the kinematic 
regime where they both apply \cite{UnifySSA}. 
Both approaches have been applied extensively in phenomenological studies \cite{qiu,Kanazawa:2000hz,Vogelsang:2005cs,siverscompare,Ans94,Boer:2003tx,BMVY,Bacch,Kouvaris:2006zy}.

For single hadron inclusive production in hadronic collisions, the large transverse momentum $P_{h\perp}$ of the produced hadron is the only hard momentum scale. For the SSAs in such single-scale hard processes, one expects the QCD collinear factorization approach to be valid \cite{Qiu_sterman}. It attributes the observed phenomenon of SSAs to three-parton correlations inside a polarized proton,
commensurate with the twist-3 nature of the overall observable.
At the leading power in the hard scale, such partonic correlations are represented by the twist-3 quark-gluon correlation function \cite{Efremov,qiu},
\ben
T_{q, F}(x, x)=\int\frac{dy_1^- dy_2^-}{4\pi}e^{ixP^+y_1^-}
\langle P,s_T|\bar{\psi}_q(0)\gamma^+\left[ \epsilon^{s_T\sigma n\bar{n}}F_\sigma^{~ +}(y_2^-)\right] \psi_q(y_1^-)|P,s_T\rangle,
\label{Tq}
\een
and by a twist-3 tri-gluon correlation function \cite{Ji:1992eu,Kang:2008qh},
\ben
T_G(x, x)=\int\frac{dy_1^- dy_2^-}{2\pi}e^{ixP^+y_1^-}\frac{1}{xP^+}\langle P,s_T|F^+_{~~\alpha}(0)\left[ \epsilon^{s_T\sigma n\bar{n}}F_\sigma^{~ +}(y_2^-)\right] F^{\alpha+}(y_1^-)|P,s_T\rangle ,
\label{Tg}
\een
where the proper gauge links have been suppressed.  There is 
a quark-gluon correlation function, $T_{q,F}$, for each quark (anti-quark) flavor $q$ 
($\bar{q}$), and there are two independent tri-gluon correlation functions, 
$T_G^{(f)}(x, x)$ and $T_G^{(d)}(x, x)$, 
because of the fact that the color of the three gluon field strengths in Eq.~(\ref{Tg}) can be neutralized by contracting with either the antisymmetric $if^{ABC}$ or the symmetric $d^{ABC}$ tensors with color indices, $A$, $B$, and $C$
\cite{Ji:1992eu,Kang:2008qh}.

Most phenomenological studies of SSAs in the twist-3 formalism performed so 
far have concentrated on the contribution by the quark-gluon correlation functions \cite{qiu,Kanazawa:2000hz,Kouvaris:2006zy}.  The role of the tri-gluon correlation function in SSAs was first studied by Ji in the context of direct-photon production in hadronic collisions \cite{Ji:1992eu}.  Recently, two of us investigated the contribution of this correlation function to the SSA for open charm production in SIDIS \cite{Kang:2008qh}. 
The twist-3 tri-gluon correlation functions have also been discussed in the
context of spin asymmetries for di-jets (or di-hadrons) in hadronic 
scattering, calculated in the TMD framework~\cite{cedran}. Here they appear 
after weighting with the transverse-momentum imbalance of the two jets.

In this paper, we study the role of the tri-gluon correlation function in the SSA for open charm production in hadron-hadron collisions. We calculate both the quark-gluon and the tri-gluon contributions to the SSA, at leading order. 
We find that the quark-gluon correlation function, extracted from data on SSAs in pion production, leads to a very small SSA for open charm production in $pp$ collisions at RHIC energy in almost all the accessible kinematic region. That is, any sizable SSAs observed for open charm production at RHIC will be a direct evidence of a non-vanishing tri-gluon correlation inside a polarized proton. 
Within a simple model for the 
$T_G^{(f)}(x, x)$ and $T_G^{(d)}(x, x)$, 
we find a fairly large SSA for open charm production at RHIC energies.  The SSA tends to be maximal in the forward region of the polarized proton and becomes small or changes sign in the backward region.  In addition, we find that our calculated SSAs for $D$ and $\bar{D}$ mesons are the same if the tri-gluon correlation function $T_G^{(d)}$ vanishes.  If $T_G^{(d)}$ is finite, the SSAs for 
$D$  and $\bar{D}$ meson production could be very different and could in fact 
be used to disentangle the contributions from the two different tri-gluon correlation functions. With data on SSAs for open charm production becoming available from RHIC~\cite{liu}, we will be able to extract the tri-gluon correlation 
functions and to learn about 
the dynamics of quantum correlations of gluons inside a polarized proton.
With additional data on the SSAs for other $pp$ processes~\cite{SSA-rhic} 
and perhaps in the
future for SSAs for $D$  and $\bar{D}$ meson production in SIDIS, a global 
analysis of QCD dynamics beyond leading twist may become possible in the 
not too distant future.

We note that predictions for the SSA in $D$ meson production at RHIC have
also been obtained within the TMD approach~\cite{anselminoD}, 
assuming that a TMD factorization holds for the single-inclusive observable
$pp\to DX$. The asymmetry was considered as a means to learn about
the gluon Sivers function. In our opinion, however, 
the twist-3 approach provides the relevant and more appropriate 
framework here, as we described above. We emphasize again that our
approach naturally involves two separate tri-gluon correlation 
functions $T_G^{(f)}(x, x)$ and $T_G^{(d)}(x, x)$, whereas only one
Sivers function occurs in the study of~\cite{anselminoD} (see 
also~\cite{muldrod}). This is expected in general to lead to discernible 
differences also in phenomenological applications. To give one
example, using a single gluon Sivers function, one will find 
the SSAs for $D$ and $\bar{D}$ meson production to be equal, whereas
in our approach they could potentially be very different, as we shall
see below. We will return to these issues in our concluding 
remarks. 

We also note that the recent work~\cite{Yuan:2008it} considers
the SSA in charm and anti-charm production at RHIC, through the
twist-3 mechanism. In this paper, only the contribution by the 
quark-gluon correlation function was considered, and the fragmentation
of the charm quark into a charmed meson was neglected.

The rest of our paper is organized as follows. In Sec.~\ref{ssa_cal}, 
we present our calculation of the SSAs for open charm production in 
hadronic collisions in terms of the QCD collinear factorization approach. 
In Sec.~\ref{ssa_phe}, we propose simple models for the tri-gluon 
correlation functions, and present our numerical estimates for the SSAs 
for open charm production in $pp$ collisions at RHIC. Finally, we summarize 
our results in Sec.~\ref{ssa_con}.

\section{Calculation of Single Transverse-Spin Asymmetry}
\label{ssa_cal}

We consider inclusive single charm meson production in a scattering process between a polarized proton $A$ of momentum $P$ and transverse spin vector $s_T$ and an unpolarized proton $B$ of momentum $P'$,
\ben
A(P,s_T)+B(P')\to h(P_h)+X,
\een
where $h$ represents the observed open charm ($D$ or $\bar{D}$) meson with momentum $P_h$ and mass $m_h$.

The spin-averaged and spin-dependent cross sections $\sigma(P_h)$, $\Delta\sigma(P_h,s_T)$ are defined as
\ben
\sigma(P_h)\equiv\frac{1}{2}\left[\sigma(P_h,s_T)+\sigma(P_h,-s_T)\right],
\qquad
\Delta\sigma(P_h,s_T)\equiv\frac{1}{2}\left[\sigma(P_h,s_T)-\sigma(P_h,-s_T)\right].
\een
The single transverse-spin asymmetry $A_N$ is defined as the ratio of $\Delta\sigma(P_h,s_T)$ and $\sigma(P_h)$,
\ben
A_N=E_{P_h}\frac{d\Delta\sigma(P_h,s_T)}{d^3P_h} \left/
E_{P_h}\frac{d\sigma(P_h)}{{d^3P_h}} \right. 
\, ,
\label{AN}
\een
for the single hadron differential cross sections.

The spin-averaged differential cross section for $D$ meson production at large transverse momentum, $P_{h\perp} > m_h$, can be written in the following 
factorized form \cite{CSS-fac,NQS-hq}:
\ben
E_{P_h}\frac{d\sigma}{d^3P_h}=\frac{ \alpha_s^2}{S}\sum_{a,b}\int \frac{dz}{z^2} D_{c\to h}(z)\int \frac{dx'}{x'}\phi_{b/B}(x')\int\frac{dx}{x}\phi_{a/A}(x)\delta\left(\tilde{s}+\tilde{t}+\tilde{u}\right)
H^{U}_{ab\to c}(\tilde{s},\tilde{t},\tilde{u}),
\label{spinavg}
\een
where $\sum_{a,b}$ represents the sum over all parton flavors and $S=(P+P')^2$ is the total collision energy squared. $\phi_{a/A}(x)$ and $\phi_{b/B}(x')$ are the standard parton distribution functions, and $D_{c\to h}(z)$ is the fragmentation function for a charm quark $c$ fragmenting into a $D$ meson. We have
neglected all dependence on the factorization and renormalization scales 
in~(\ref{spinavg}). 
When $P_{h\perp} \gg m_h$, a sizable amount of charm mesons could be 
produced from the fragmentation of a gluon or a light quark or 
anti-quark.  Then, the sum $\sum_{a,b}$ in Eq.~(\ref{spinavg}) would need to 
be complemented by an additional sum over the fragmenting parton, 
running over all possible light parton flavors as well as the charm.  
However, we do not expect a large fragmentation contribution from light 
partons to charm meson production at the current RHIC energies, and we will 
limit our calculations to the fragmentation of charm and anti-charm 
quarks in this paper. 

In Eq.~(\ref{spinavg}), $H^{U}_{ab\to c}$ is a short-distance hard part for two partons of flavor $a$ and $b$ to produce a charm quark $c$. At the lowest order, it gets contributions from the light quark-antiquark annihilation and gluon-gluon fusion subprocesses, as sketched in Fig.~\ref{unlow}, and is given by
\ben
H^U_{q\bar{q}\to c}&=&\frac{C_F}{N_C}\left[\frac{\tilde{t}^2+\tilde{u}^2+2m_c^2\tilde{s}}{\tilde{s}^2}\right],\nnu
H^U_{gg\to c}&=&\frac{1}{2N_C}\left[\frac{1}{\tilde{t}\tilde{u}}-\frac{N_C}{C_F}\frac{1}{\tilde{s}^2}\right]
\left[\tilde{t}^2+\tilde{u}^2+4m_c^2\tilde{s}-\frac{4m_c^4\tilde{s}^2}{\tilde{t}\tilde{u}}\right],
\een
where $\tilde{s},\tilde{t},\tilde{u}$ are defined at the partonic level as
\ben
\tilde{s}=(xP+x'P')^2,
\qquad
\tilde{t}=(xP-p_c)^2-m_c^2,
\qquad
\tilde{u}=(x'P'-p_c)^2-m_c^2,
\een
with $p_c$ and $m_c$ the momentum and mass of the charm quark that fragments into the $D$ meson, respectively.
\bef
\psfig{file=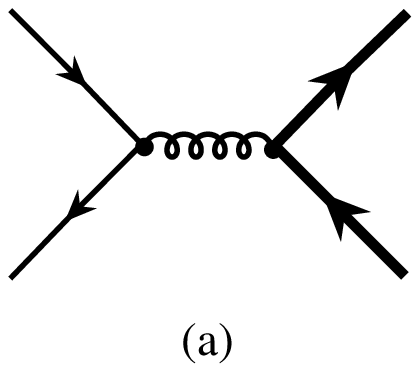,width=1.0in}
\hskip 0.5in
\psfig{file=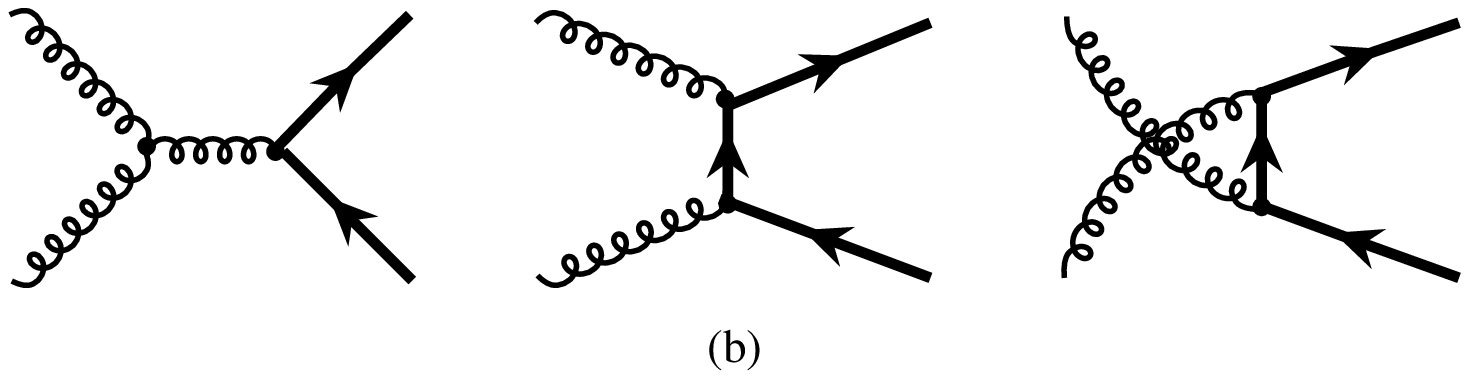,width=3.7in}
\caption{Lowest order Feynman diagram for light quark-antiquark annihilation (a) and for gluon-gluon fusion to a pair of heavy quarks.}
\label{unlow}
\eef

The transverse spin-dependent cross section also gets contributions from both the $q\bar{q}$ annihilation and $gg$ fusion channels. Following the generalized factorization theorem \cite{Qiu_sterman}, the cross section in the 
$q\bar{q}$ annihilation channel can be expressed in terms of a general twist-3 quark-gluon correlation function \cite{Kouvaris:2006zy},
\ben
d\Delta\sigma(s_T)\propto \frac{1}{2 S}\sum_{q}\int dz D_{c\to h}(z) 
\int \frac{dx'}{x'}\phi_{\bar{q}/B}(x')
\int dx_1 dx_2 T_{q,F}(x_1,x_2)\ i \epsilon^{\rho s_T n\bar{n}}
\lim_{k_\perp\to 0}\frac{\partial}{\partial k_\perp^\rho}H_{q\bar{q}\to c}(x_1,x_2,k_\perp),
\label{Dqqtoc}
\een
with the quark-gluon correlation function $T_{q,F}(x_1, x_2)$ given by
\ben
T_{q, F}(x_1, x_2)=\int\frac{dy_1^- dy_2^-}{4\pi}e^{ix_1P^+y_1^- + i(x_2-x_1)P^+y_2^-}
\langle P,s_T|\bar{\psi}_q(0)\gamma^+\left[ \epsilon^{s_T\sigma n\bar{n}}F_\sigma^{~ +}(y_2^-)\right] \psi_q(y_1^-)|P,s_T\rangle\, ,
\een
whose diagonal term, at $x_1=x_2$, is equal to the correlation function in Eq.~(\ref{Tq}).
In order to generate a non-vanishing SSA, a strong interaction phase is necessary, which comes from the interference between a real part of the
scattering amplitude and an imaginary part of the partonic scattering 
amplitude with an extra gluon, as shown in Fig.~\ref{qqspin}. Technically, 
the imaginary part arises when the virtual momentum integral of the extra 
gluon is evaluated by the residue of an unpinched pole from a propagator in 
the amplitude with an extra gluon. Such a propagator is indicated by the 
short bars in the diagrams in Fig.~\ref{qqspin}. The phase can arise from the attachment of the extra gluon to either the initial-state parton, or the final-state charm quark, which we will refer to as initial-state and final-state interactions, respectively.
\bef
\psfig{file=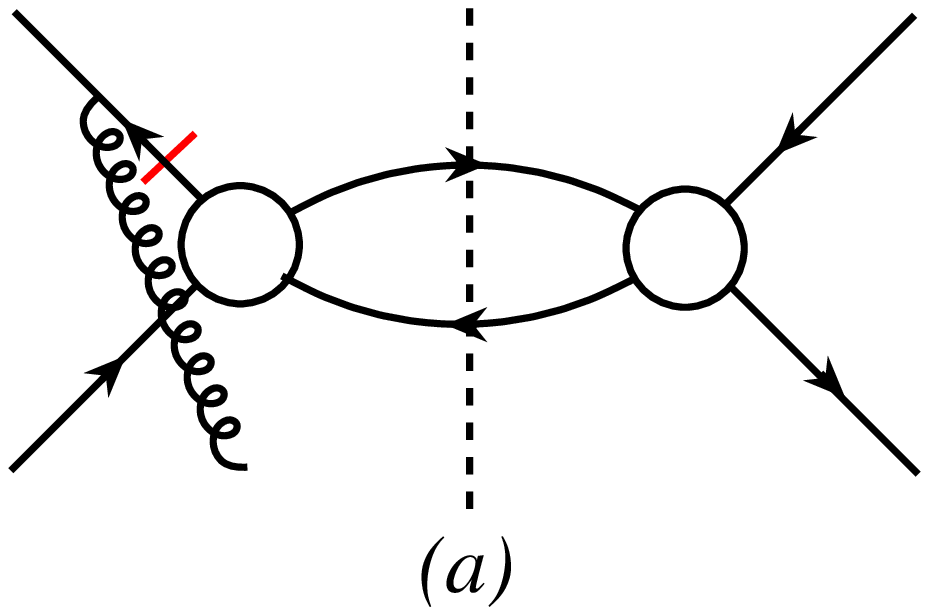,width=1.35in}\hskip 0.25in
\psfig{file=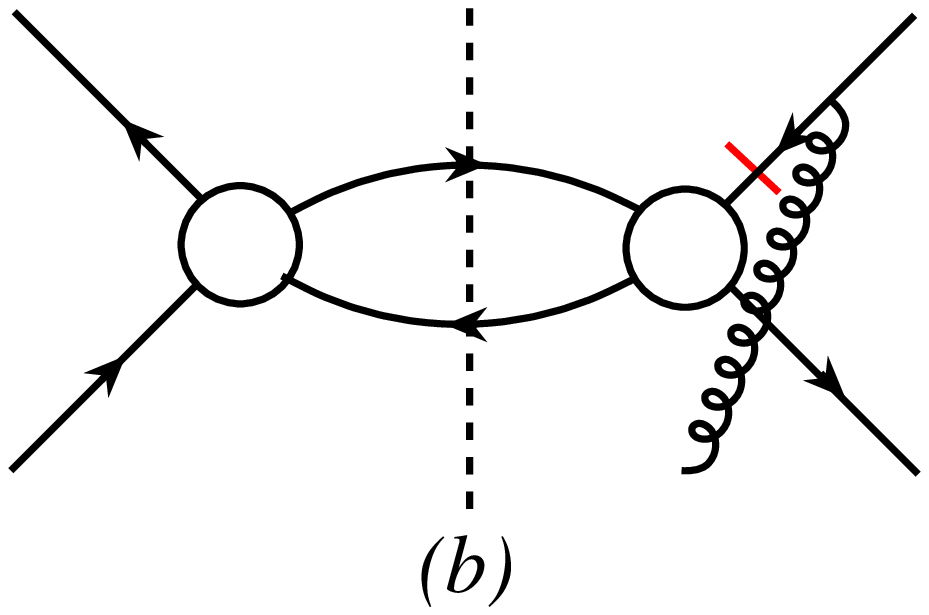,width=1.35in}\hskip 0.25in
\psfig{file=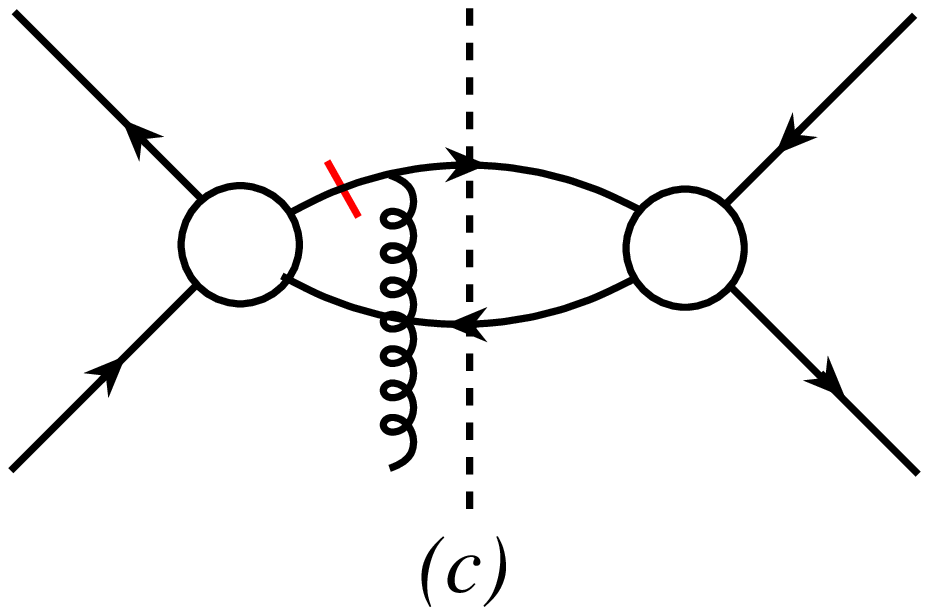,width=1.35in}\hskip 0.25in
\psfig{file=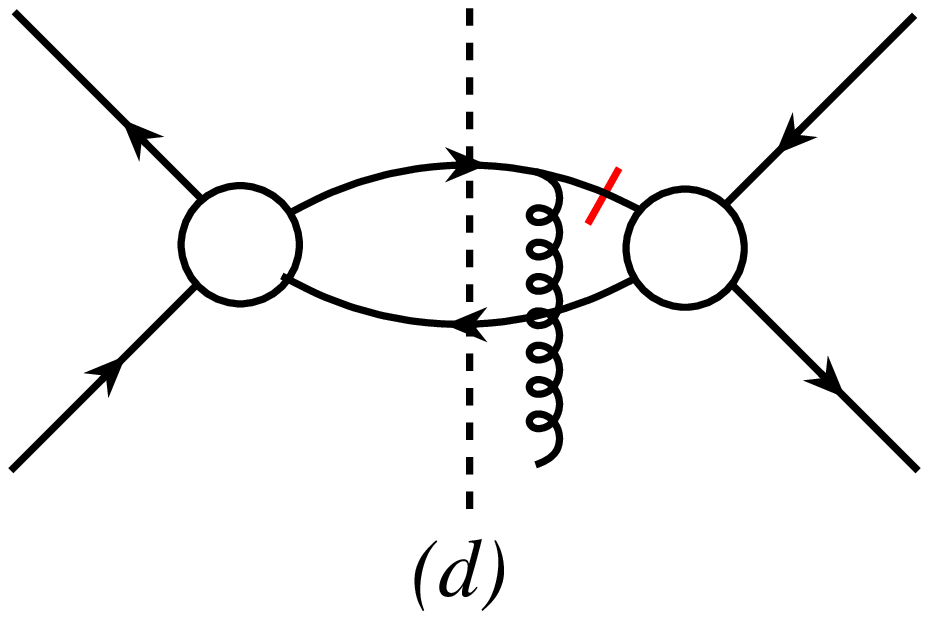,width=1.35in}
\caption{Feynman diagrams that give the twist-3 contribution to the spin-dependent cross section in the quark-antiquark annihilation channel: initial-state interaction (a), (b), and final-state interaction (c), (d). The short bar indicates the propagator that produces the unpinched pole.}
\label{qqspin}
\eef

At lowest order, there are four diagrams contributing to the twist-3 
polarized cross section in the quark-antiquark annihilation channel, as sketched in Fig.~\ref{qqspin}, where the blob is given by the diagram 
in Fig.~\ref{unlow}$(a)$. Using the techniques well established in literature (see, for example, \cite{Kouvaris:2006zy}), we have
\ben
\left.
E_{P_h}\frac{d\Delta\sigma}{d^3P_h}\right|_{q\bar{q}\to c\bar{c}}&=&\frac{ \alpha_s^2}{S}\sum_{q}\int \frac{dz}{z^2} D_{c\to h}(z)\int \frac{dx'}{x'}\phi_{\bar{q}/B}(x')\int\frac{dx}{x}\sqrt{4\pi\alpha_s}\left(\frac{\epsilon^{P_h s_T n \bar{n}}}{z\tilde{u}}\right)\delta\left(\tilde{s}+\tilde{t}+\tilde{u}\right)\nnu
&\times&
\left[\left(T_{q, F}(x,x)-x\frac{d}{dx}T_{q, F}(x,x)\right)H_{q\bar{q}\to c}(\tilde{s},\tilde{t},\tilde{u})
+T_{q, F}(x,x){\cal H}_{q\bar{q}\to c}(\tilde{s},\tilde{t},\tilde{u})\right],
\label{qqtoD}
\een
where $H_{q\bar{q}\to c}$ can be written as
\ben
H_{q\bar{q}\to c}=H_{q\bar{q}\to c}^{I}+H_{q\bar{q}\to c}^{F}\left(1+\frac{\tilde{u}}{\tilde{t}}\right),
\een
and likewise for ${\cal H}_{q\bar{q}\to c}$, and where the
corresponding hard parts are given by
\ben
H^I_{q\bar{q}\to c}&=&\frac{1}{2N_C^2}\left[\frac{\tilde{t}^2+\tilde{u}^2+2m_c^2\tilde{s}}{\tilde{s}^2}\right],
\nnu
H^F_{q\bar{q}\to c}&=&\frac{N_C^2-2}{2N_C^2}\left[\frac{\tilde{t}^2+\tilde{u}^2+2m_c^2\tilde{s}}{\tilde{s}^2}\right],
\\
{\cal H}^I_{q\bar{q}\to c}&=&\frac{1}{2N_C^2}\left[\frac{2m_c^2}{\tilde{s}}\right],
\nnu
{\cal H}^F_{q\bar{q}\to c}&=&\frac{N_C^2-2}{2N_C^2}\left[\frac{2m_c^2}{\tilde{s}}\right]\, .
\een
Note that ${\cal H}^I_{q\bar{q}\to c}$ and ${\cal H}^F_{q\bar{q}\to c}$
are proportional to the charm quark mass.  As a check of our results,
when $m_c^2\to 0$ the spin-dependent cross section in Eq.~(\ref{qqtoD}) becomes
identical to the one for pion production through the $q\bar{q}
\to q'\bar{q}'$ 
channel \cite{Kouvaris:2006zy} (if one replaces the $D$ meson fragmentation 
function by the pion fragmentation function).

The spin-dependent cross section for $\bar{D}$ meson production can be calculated in the same way. The Feynman diagrams are the same as those for $D$ meson production in Fig.~\ref{qqspin}, except that the extra gluon should be attached to the anti-charm $\bar{c}$ quark for the final-state interaction. The cross section for $\bar{D}$ meson production 
has the same factorized form as that in Eq.~(\ref{qqtoD}), 
with the fragmentation function $D_{c\to D}(z)$ replaced by $D_{\bar{c}
\to \bar{D}}(z)$, and the hard parts given by
\ben
H^I_{q\bar{q}\to \bar{c}}&=&\frac{1}{2N_C^2}\left[\frac{\tilde{t}^2+\tilde{u}^2+2m_c^2\tilde{s}}{\tilde{s}^2}\right],
\nnu
H^F_{q\bar{q}\to \bar{c}}&=&\frac{1}{N_C^2}\left[\frac{\tilde{t}^2+\tilde{u}^2+2m_c^2\tilde{s}}{\tilde{s}^2}\right],
\\
{\cal H}^I_{q\bar{q}\to \bar{c}}&=&\frac{1}{2N_C^2}\left[\frac{2m_c^2}{\tilde{s}}\right],
\nnu
{\cal H}^F_{q\bar{q}\to \bar{c}}&=&\frac{1}{N_C^2}\left[\frac{2m_c^2}{\tilde{s}}\right].
\een
These short-distance hard parts are consistent with those presented 
in the calculation of the SSAs for heavy quark and anti-quark production 
in hadronic collisions \cite{Yuan:2008it}. We note that the 
hard parts for $\bar{q}q$ scattering are obtained from those for
$q\bar{q}$ by $H^{I,F}_{\bar{q}q\to c}=-H^{I,F}_{q\bar{q}\to \bar{c}}$
and $H^{I,F}_{\bar{q}q\to \bar{c}}=-H^{I,F}_{q\bar{q}\to c}$, and
likewise for the ${\cal H}^{I,F}$.

Similar to Eq.~(\ref{Dqqtoc}), the spin-dependent cross section for the
$gg$ fusion channel has the following factorized form:
\ben
d\Delta\sigma(s_T)\propto \frac{1}{2 S}\int dz D_{c\to h}(z) \int \frac{dx'}{x'}\phi_{g/B}(x')
\int dx_1 dx_2 \widetilde{T}_G(x_1,x_2)\ i \epsilon^{\rho s_T n\bar{n}}
\lim_{k_\perp\to 0}\frac{\partial}{\partial k_\perp^\rho}H_{gg\to c}(x_1,x_2,k_\perp),
\label{Dggtoc}
\een
where $\widetilde{T}_G(x_1, x_2)$ is defined as
\ben
\widetilde{T}_G(x_1,x_2)=\int \frac{P^+dy_1^- dy_2^-}{2\pi}e^{ix_1P^+y_1^- + i (x_2-x_1)P^+y_2^-}d_{\alpha\beta}\langle P,s_T| A^{\alpha}(0)\left[ \epsilon^{s_T\sigma n\bar{n}}F_\sigma^{~ +}(y_2^-)\right] A^{\beta}(y_1^-)|P, s_T\rangle,
\een
with $d_{\alpha\beta}=-g_{\alpha\beta}+\bar{n}_\alpha n_\beta+\bar{n}_\beta n_\alpha$. $\widetilde{T}_G(x_1,x_2)$ is related to the tri-gluon correlation function through $T_G(x,x)=x\widetilde{T}_G(x,x)$.
\bef
\psfig{file=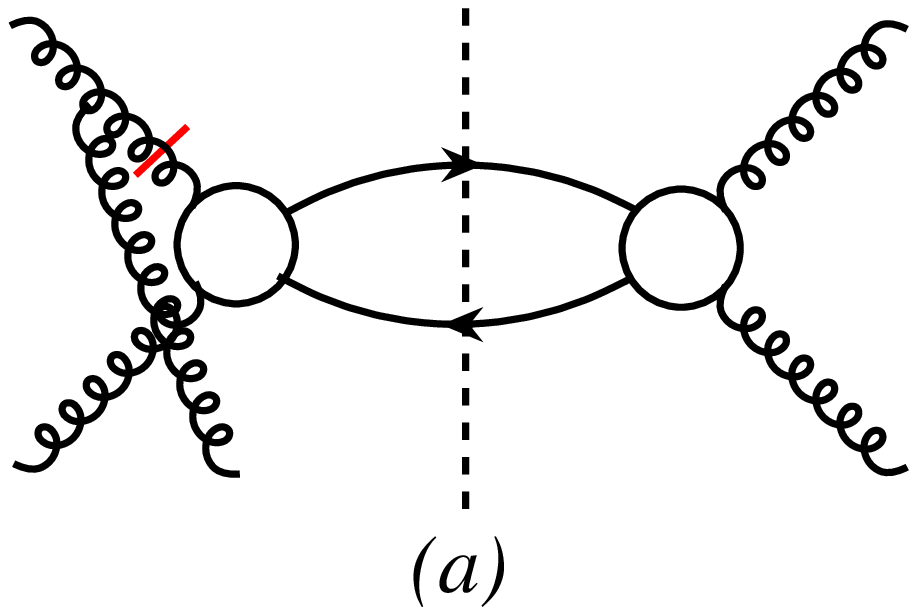,width=1.35in}\hskip 0.25in
\psfig{file=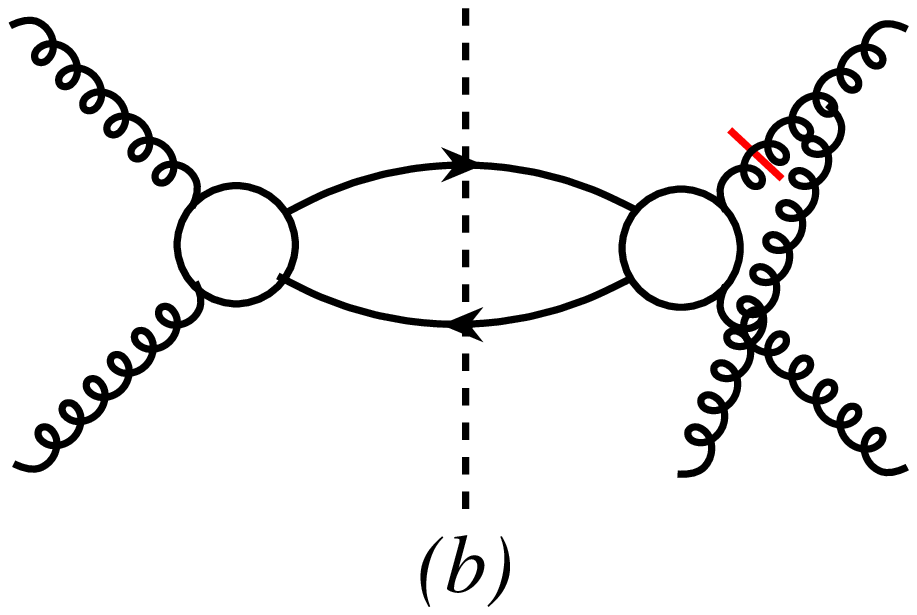,width=1.35in}\hskip 0.25in
\psfig{file=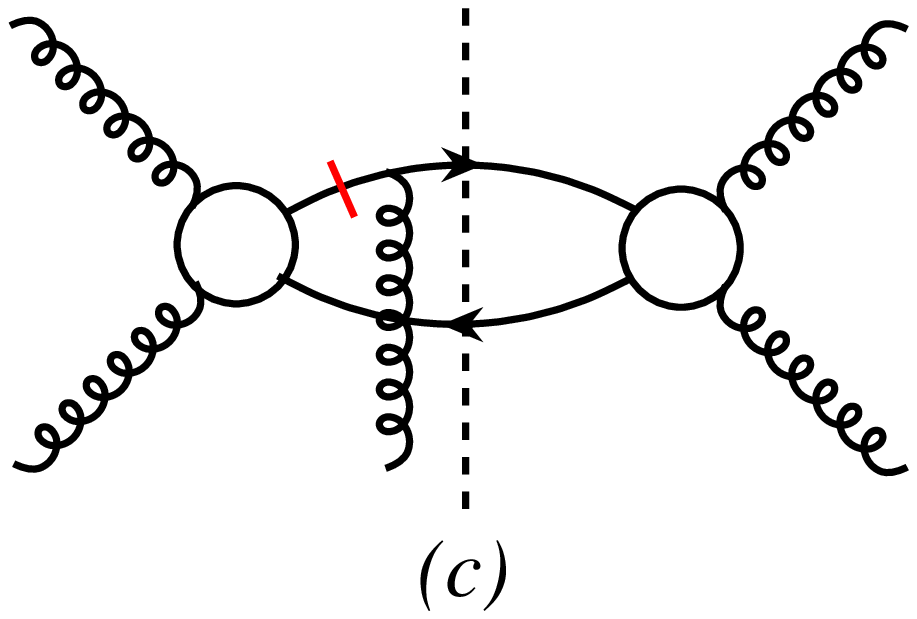,width=1.35in}\hskip 0.25in
\psfig{file=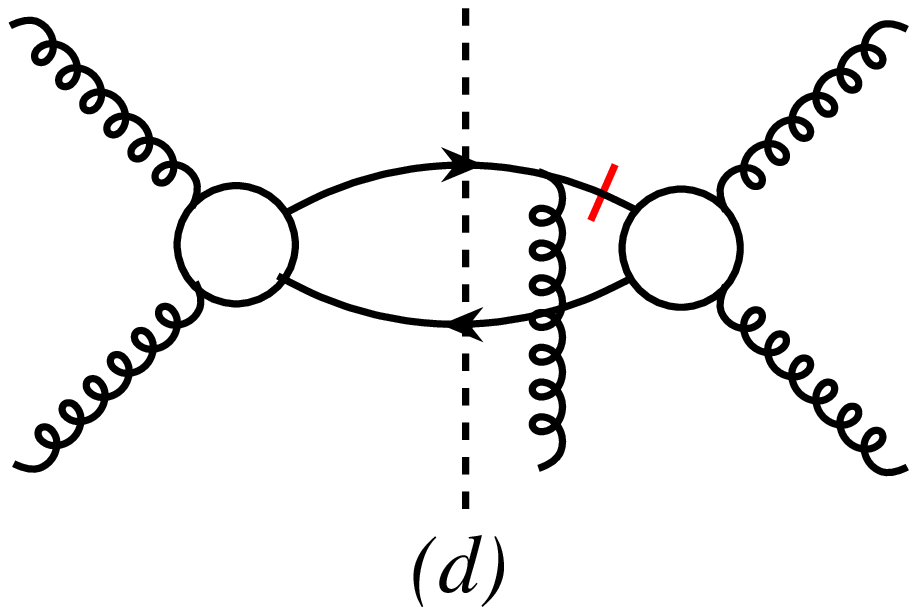,width=1.35in}
\caption{Feynman diagrams that give the twist-3 contribution to the spin-dependent cross section in the gluon-gluon fusion channel: 
initial-state interaction (a), (b), and final-state interaction (c), (d). 
The short bar indicates the propagator that produces the pole.}
\label{ggspin}
\eef

To calculate the partonic hard part, $H_{gg\to c}$, in Eq.~(\ref{Dggtoc}), we need to consider Feynman diagrams with either initial-state or final-state interactions, as sketched in Fig.~\ref{ggspin}, where the blob is given by the 
sum of the three diagrams in Fig.~\ref{unlow}$(b)$. Hence, each diagram in Fig.~\ref{ggspin} corresponds to nine diagrams.  Instead of four diagrams in Fig.~\ref{qqspin} for the quark-antiquark annihilation subprocess, we have a total of 36 diagrams for gluon-gluon fusion.  By evaluating these diagrams, we obtain the contribution to the spin-dependent cross section,
\ben
\left.
E_{P_h}\frac{d\Delta\sigma}{d^3P_h}\right|_{gg\to c\bar{c}}&=&\frac{ \alpha_s^2}{S}\sum_{i=f,d}\int \frac{dz}{z^2} D_{c\to h}(z)\int \frac{dx'}{x'}\phi_{g/B}(x')\int \frac{dx}{x}
\sqrt{4\pi\alpha_s}\left(\frac{\epsilon^{P_h s_T n \bar{n}}}{z\tilde{u}}\right)\delta\left(\tilde{s}+\tilde{t}+\tilde{u}\right)\nnu
&\times&
\left[\left(T^{(i)}_{G}(x,x)-x\frac{d}{dx}T^{(i)}_{G}(x,x)\right)H^{(i)}_{gg\to c}(\tilde{s},\tilde{t},\tilde{u})
+T^{(i)}_{G}(x,x){\cal H}^{(i)}_{gg\to c}(\tilde{s},\tilde{t},\tilde{u})\right],
\label{ggtoD}
\een
where the sum, $\sum_{i=f,d}$, is over the two correlation functions $T_G^{(f)}(x,x)$ and $T_G^{(d)}(x,x)$. The partonic hard
part $H^{(i)}_{gg\to c}$ can be written as 
\ben
H^{(i)}_{gg\to c}=H_{gg\to c}^{I(i)}+H_{gg\to c}^{F(i)}\left(1+\frac{\tilde{u}}{\tilde{t}}\right),
\een
and likewise for ${\cal H}^{(i)}_{gg\to c}$, and we find
\ben
H^{I(f)}_{gg\to c}&=&-\frac{1}{8C_F}\frac{\tilde{t}^2+\tilde{u}^2}{\tilde{t}\tilde{u}\tilde{s}^2}
\left[\tilde{t}^2+\tilde{u}^2+4m_c^2\tilde{s}-\frac{4m_c^4\tilde{s}^2}{\tilde{t}\tilde{u}}\right],
\nnu
H^{I(d)}_{gg\to c}&=&-\frac{1}{8C_F}\frac{\tilde{u}-\tilde{t}}{\tilde{t}\tilde{u}\tilde{s}}
\left[\tilde{t}^2+\tilde{u}^2+4m_c^2\tilde{s}-\frac{4m_c^4\tilde{s}^2}{\tilde{t}\tilde{u}}\right],
\\
H^{F(f)}_{gg\to c}&=&H^{F(d)}_{gg\to c}=\left[
\frac{N_C}{4\left(N_C^2-1\right)}\frac{\tilde{u}}{\tilde{t}\tilde{s}^2}
-\frac{1}{4N_C\left(N_C^2-1\right)}\frac{1}{\tilde{t}\tilde{u}}\right]\left[\tilde{t}^2+\tilde{u}^2+4m_c^2\tilde{s}-\frac{4m_c^4\tilde{s}^2}{\tilde{t}\tilde{u}}\right],
\een
and
\ben
{\cal H}^{I(f)}_{gg\to c}&=&-\frac{1}{2C_F}\frac{m_c^2\left(\tilde{t}^2+\tilde{u}^2\right)\left(\tilde{t}\tilde{u}-2m_c^2\tilde{s}\right)}{\tilde{s}\tilde{t}^2\tilde{u}^2},
\nnu
{\cal H}^{I(d)}_{gg\to c}&=&-\frac{1}{2C_F}\frac{m_c^2\left(\tilde{u}-\tilde{t}\right)\left(\tilde{t}\tilde{u}-2m_c^2\tilde{s}\right)}{\tilde{t}^2\tilde{u}^2},
\nnu
{\cal H}^{F(f)}_{gg\to c}&=&{\cal H}^{F(d)}_{gg\to c}=-\left[\frac{1}{N_C(N_C^2-1)}\frac{1}{\tilde{u}^2}-\frac{N_C}{N_C^2-1}\frac{1}{\tilde{s}^2}\right]
\frac{m_c^2\tilde{s}(\tilde{t}\tilde{u}-2m_c^2\tilde{s})}{\tilde{t}^2}.
\een

The gluon-gluon subprocess of course 
also contributes to the cross section for $\bar{D}$ meson production.  
The corresponding partonic hard parts for producing an anti-charm 
quark are given by
\ben
H^{(f)}_{gg\to \bar{c}}=H^{(f)}_{gg\to c},
\qquad
H^{(d)}_{gg\to \bar{c}}=-H^{(d)}_{gg\to c},
\nnu
{\cal H}^{(f)}_{gg\to \bar{c}}={\cal H}^{(f)}_{gg\to c}\, ,
\qquad
{\cal H}^{(d)}_{gg\to \bar{c}}=-{\cal H}^{(d)}_{gg\to c}\, ,
\label{DtoDbar}
\een
where the sign difference of the partonic hard parts for the
$T_G^{(d)}$ contribution 
will be responsible for the difference of the SSAs for $D$  and 
$\bar{D}$ meson production that will be discussed in the next section.
We note that this sign difference can also be observed in the 
expressions for the ``gluonic pole matrix elements'' given 
in~\cite{cedran}.

We point out that the compact dependence of the spin-dependent cross 
section on the combinations $T(x,x)-xT'(x,x)$ of the twist-3 correlation 
functions found in Ref.~\cite{Kouvaris:2006zy} for the
``massless'' case of pion production in hadronic collisions, 
is violated for the production of $D$ (or $\bar{D}$) mesons 
by the additional non-derivative terms in Eqs.~(\ref{qqtoD}) and 
(\ref{ggtoD}). The violation is caused by the heavy quark mass since 
the additional terms vanish when $m_c\to 0$. In fact,
we observe that the hard parts we have derived satisfy the 
following relation:
\ben
{\cal H}^{I,F}_{ab\to c}=m_c^2\frac{dH^{I,F}_{ab\to c}}{dm_c^2} \; , 
\een
separately for any of the various contributions considered above
(and likewise for $\bar{c}$ production). This
connection is likely a consequence of the ``master formula'' for
twist-3 soft-gluon-pole contributions derived in~\cite{koiketanaka}.

We also note that ``soft-fermion pole'' contributions~\cite{qiu}, 
for which the pole in the hard-scattering function is taken in such 
a way that the initial quark, rather than the initial gluon, 
becomes soft, are absent for the $q\bar{q}$ process at the leading
order. This is because 
$q\bar{q}$ annihilation proceeds through an $s$-channel diagram,
whereas soft-fermion poles would only appear in $t$-channel diagrams.
If they were present, such contributions would involve the function 
$T_{q, F}(0, x)$. For the tri-gluon correlation contribution, terms 
proportional to $T_G^{(f)}(0, x)$ and $T_G^{(d)}(0, x)$ are automatically 
included in our calculations, thanks to the symmetry properties of the
correlation functions \cite{Ji:1992eu}.

Combining the factorized cross sections in Eqs.~(\ref{qqtoD}) and (\ref{ggtoD}) with the corresponding partonic hard parts, we have 
\ben
E_{P_h}\frac{d\Delta\sigma}{d^3P_h}=
\left. E_{P_h}\frac{d\Delta\sigma}{d^3P_h}\right|_{q\bar{q}\to c\bar{c}}
+
\left. E_{P_h}\frac{d\Delta\sigma}{d^3P_h}\right|_{gg\to c\bar{c}}
\label{spindep}
\een
for the leading-order contribution to the transverse-spin-dependent cross section for $D$ (or $\bar{D}$) meson production in hadronic collisions. The 
corresponding single transverse-spin asymmetry is obtained by 
substituting Eqs.~(\ref{spinavg}) and (\ref{spindep}) into Eq.~(\ref{AN}).

\section{Phenomenology}
\label{ssa_phe}
We now use the expressions we have derived in the previous section
to obtain phenomenological
results suitable for RHIC. We will propose simple models 
for the tri-gluon correlation functions $T_G^{(f)}(x,x)$ and $T_G^{(d)}(x,x)$, and then estimate the size of the SSA for $D$  and $\bar{D}$ meson production in $p^\uparrow p$ 
collisions at RHIC at $\sqrt{s}=200$~GeV. 

In our studies, we adopt the CTEQ6L parton distribution functions 
\cite{Pumplin:2002vw} for the unpolarized proton. We use the 
charm-to-$D$ meson fragmentation 
function of Ref.~\cite{Kneesch:2007ey}. In order to treat the
kinematic charm mass effects in the fragmentation process, 
we adopt one of the choices introduced in Ref. \cite{Cacciari:2003uh}, 
which corresponds to setting $P_{h\perp}=zp_{c\perp}$ and $y_D=y_c\equiv y$, 
where $P_{h\perp}$ ($p_{c\perp}$) and $y_D$ ($y_c$) are the transverse 
momentum and rapidity of the $D$ meson (charm quark), respectively.
With this choice, we then have for $\tilde{s}, \tilde{t}, \tilde{u}$ 
and the Feynman variable $x_F$:
\ben
\tilde{s}=x'xS,
\qquad
\tilde{t}=-x m_{c\perp}\sqrt{S}e^{-y},
\qquad
\tilde{u}=-x' m_{c\perp}\sqrt{S}e^{y},
\qquad
x_F=\frac{m_{h\perp}}{\sqrt{S}}\left(e^y-e^{-y}\right),
\een
where $m_{c\perp}=\sqrt{m_c^2+p_{c\perp}^2}$ and $m_{h\perp}=
\sqrt{m_h^2+P_{h\perp}^2}$.
We further assume that $D_{\bar{c}\to\bar{D}}(z)=D_{c\to D}(z)$ for
the $\bar{D}$ meson fragmentation functions. We choose the factorization 
scale to be equal to the renormalization scale throughout, and set 
$\mu=\sqrt{m_c^2+P_{h\perp}^2}$ with $m_c=1.3$ GeV.

In order to calculate numerical values for the SSAs for the production of $D$ 
or $\bar{D}$ mesons, we need the unknown, but universal, tri-gluon 
correlation functions $T_G^{(f,d)}(x,x)$.  Because of their nonperturbative 
nature, both the quark-gluon correlation functions $T_{q,F}(x,x)$ and the
tri-gluon correlation functions $T_G^{(f,d)}(x,x)$ should in principle 
be extracted from experimental data on SSAs. For the $T_{q,F}(x,x)$, this 
was done in Ref.~\cite{Kouvaris:2006zy} by fits to the experimental 
data \cite{SSA-fixed-tgt,SSA-rhic} for the SSA in $pp\to \pi X$. We adopt
in the following the set referred to as ``Fit II'' in \cite{Kouvaris:2006zy}.
We note that the tri-gluon correlations were not taken into account
in \cite{Kouvaris:2006zy}, so that no direct information on these
is available so far. For the purpose of estimating the SSAs and 
motivating future experimental measurements, we therefore follow the 
arguments given in Ref.~\cite{Kang:2008qh} and model the $T_G^{(f,d)}(x,x)$ as
\ben
T_G^{(f)}(x,x)=\lambda_f G(x),
\qquad
T_G^{(d)}(x,x)=\lambda_d G(x),
\een
with $G(x)$ the ordinary unpolarized gluon distribution function. In order 
to cover a range of possibilities for the nonperturbative correlation 
functions, we introduce three sets of values for the parameters $\lambda_f$ 
and $\lambda_d$: (1)\ $\lambda_f=\lambda_d=0.07$ GeV, (2)\ $\lambda_f=
\lambda_d=0$, and (3)\ $\lambda_f=-\lambda_d=0.07$ GeV, corresponding
to the assumptions:\ $T_G^{(f)}=T_G^{(d)}$, $T_G^{(f)}=T_G^{(d)}=0$, and 
$T_G^{(f)}=-T_G^{(d)}$, respectively. In principle, the signs and the values 
of $\lambda_f$ and $\lambda_d$, as well as the functional form of the 
correlation functions should be fixed by future data. Comparing SSAs for
physical cross sections involving the same non-perturbative twist-three 
correlation functions but different partonic hard subprocesses will
provide stringent tests of QCD dynamics and the twist-3 factorization 
we use.

\bef
\psfig{file=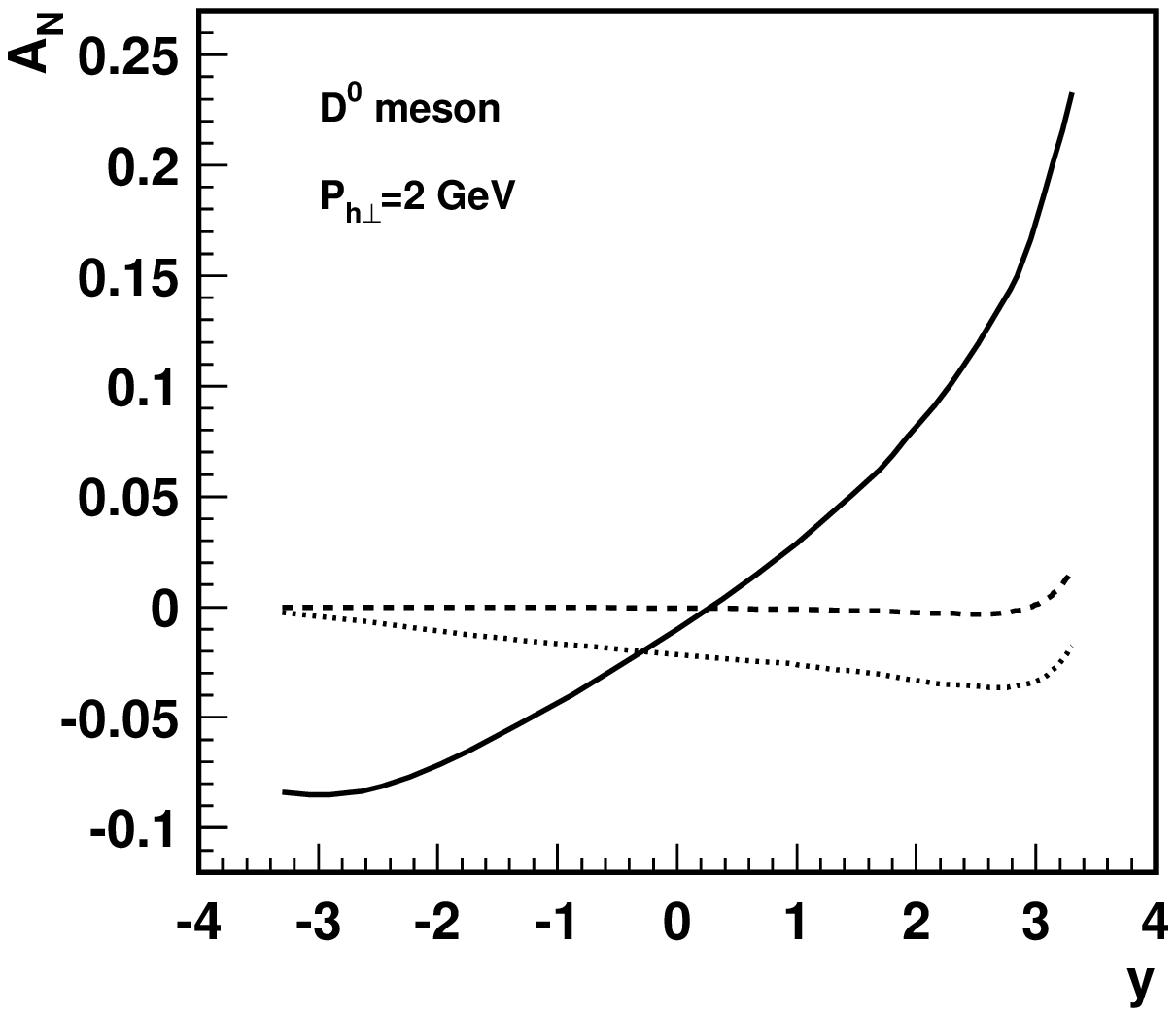,width=2.5in}
\hskip 0.2in
\psfig{file=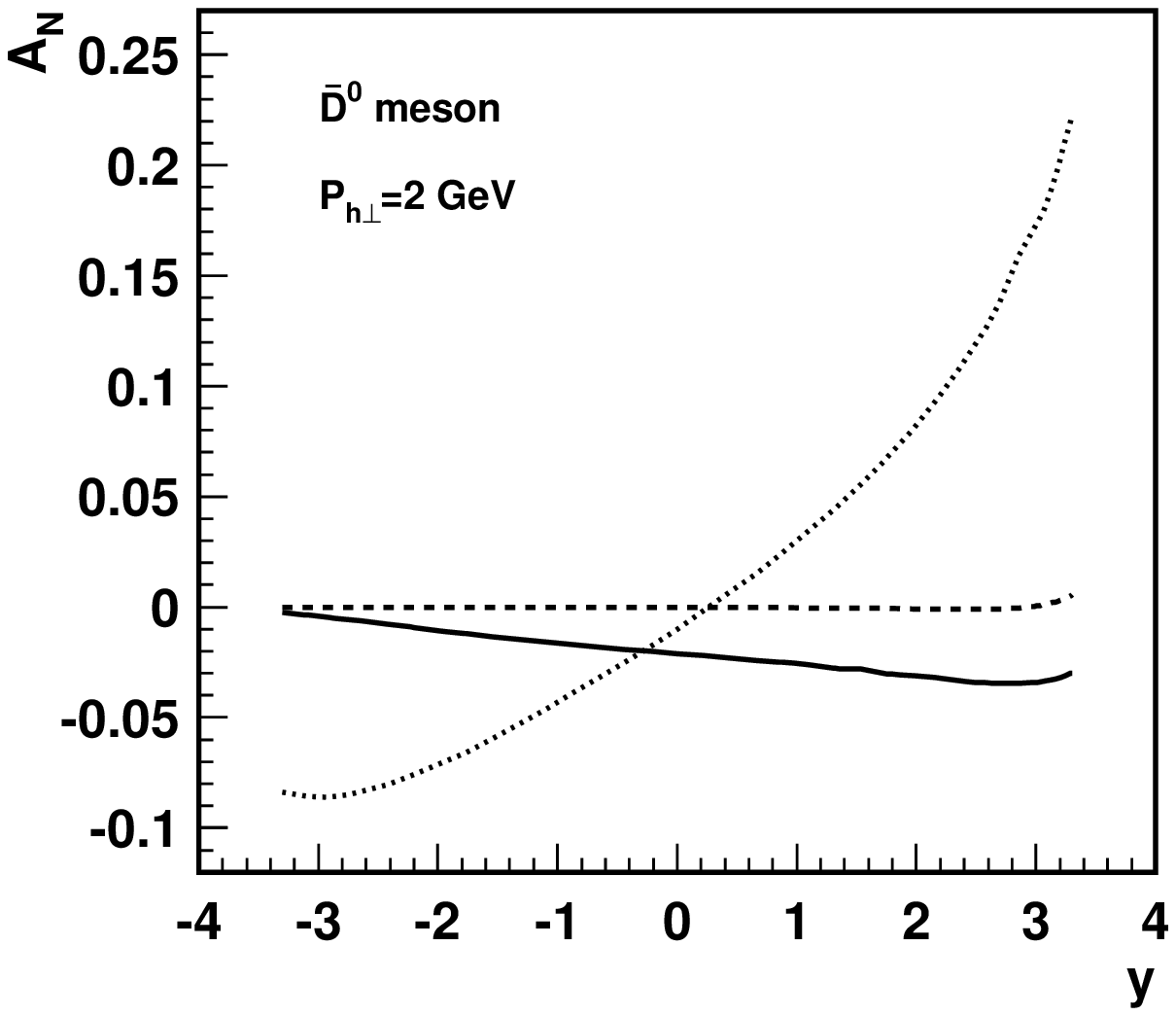,width=2.5in}
\caption{The SSA as a function of rapidity $y$ for $D^0$ meson (left) 
and $\bar{D}^0$ meson production (right) at $\sqrt{s}=200$ GeV and 
$P_{h\perp}=2$~GeV. The 
curves are: solid ($\lambda_f=\lambda_d=0.07$ GeV), dashed 
($\lambda_f=\lambda_d=0$), dotted ($\lambda_f=-\lambda_d=0.07$ GeV).}
\label{y_dep}
\eef

In Figs.~\ref{y_dep} and \ref{xf_dep} we plot the SSAs, $A_N$, 
for the production 
of $D$ and $\bar{D}$ mesons as functions of rapidity $y$ and 
Feynman-$x_F$, respectively. We count positive rapidity in the forward 
direction of the polarized proton. 
The solid, dashed, and dotted curves correspond 
to the three sets of parameters: $\lambda_f=\lambda_d=0.07$ GeV, $\lambda_f=\lambda_d=0$, and $\lambda_f=-\lambda_d=0.07$ GeV, respectively. From the dashed curves in Figs.~\ref{y_dep} and \ref{xf_dep}, it is clear that the quark-gluon correlation function $T_{q,F}$ alone generates a very small single transverse-spin asymmetry at RHIC energy.  This is because of the dominance of the 
$gg$ fusion contribution over the $q\bar{q}$ one in the spin-averaged 
cross section in the denominator of $A_N$. In other words, any significant 
size of the SSA in open charm production signals the 
discovery of tri-gluon correlations inside a polarized proton.  

\bef
\psfig{file=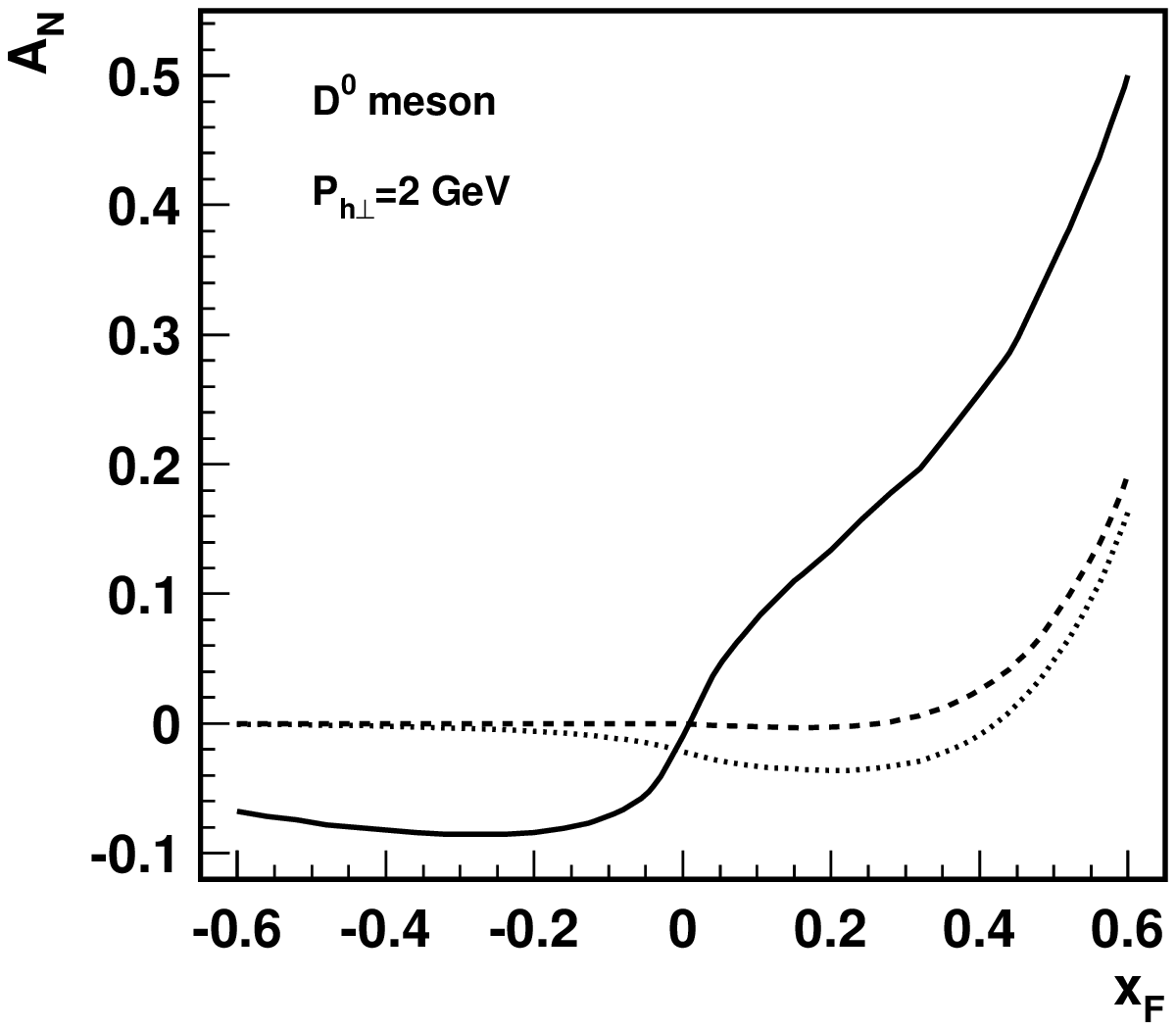,width=2.5in}
\hskip 0.2in
\psfig{file=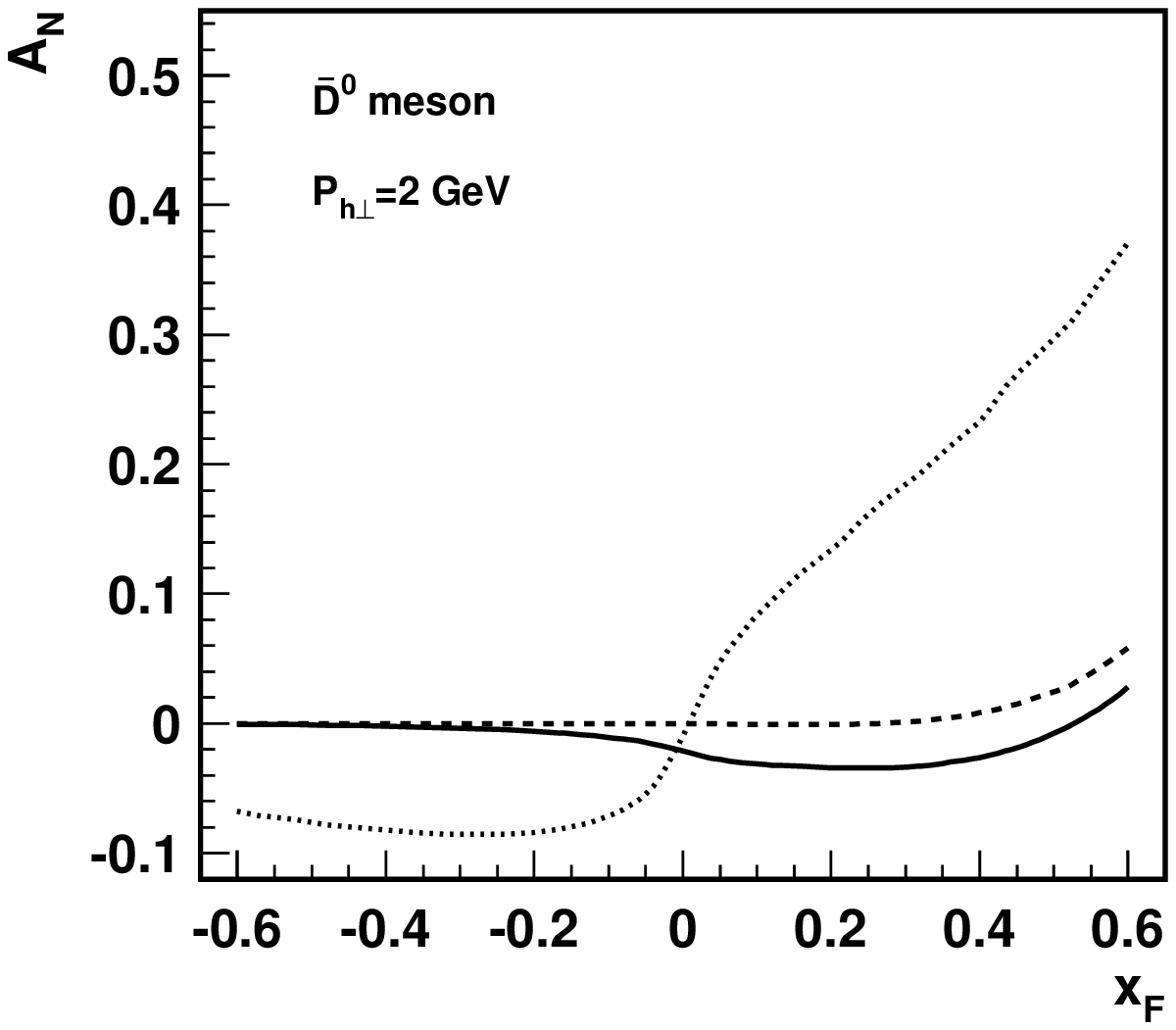,width=2.5in}
\caption{Same as Fig.~\ref{y_dep}, but as a function of Feynman-$x_F$.}
\label{xf_dep}
\eef

The difference between the solid and dotted curves in Figs.~\ref{y_dep} 
and \ref{xf_dep} indicates that the two tri-gluon correlation functions, 
$T_G^{(f)}$ and $T_G^{(d)}$, may both play very important, but different, 
roles for the SSA in $D$ and $\bar{D}$ meson production.  In the case of 
$D$ mesons, a large $A_N$ (see the solid curve) is obtained 
when $\lambda_f=\lambda_d=0.07$ GeV, i.e., when $T_G^{(f)}$ and $T_G^{(d)}$ 
have the same sign.  However, when their signs are opposite, their 
contributions to the SSA tend to cancel, leading to a much smaller SSA 
(dotted curve). On the contrary, for $\bar{D}$ meson production, the 
largest $A_N$ is found when $T_G^{(f)}$ and $T_G^{(d)}$ have opposite signs. 
This is due to the fact that, as shown in Eq.~(\ref{DtoDbar}), the partonic 
hard parts associated with $T_G^{(d)}$ change sign when going from 
charm to anti-charm production, while the hard parts for $T_G^{(f)}$ 
remain the same.  As a result, the SSA for $\bar{D}$ mesons is 
much smaller if $T_G^{(f)}$ and $T_G^{(d)}$ have the same sign.  
In addition, as seen from Figs.~\ref{y_dep} and \ref{xf_dep},
scans of the SSA from the forward to the backward region
may provide good sources of information on the $x$-dependence or the 
functional form of the tri-gluon correlation functions, in particular
if a sign change occurs. It is also striking to see that the asymmetry
for either $D$ or $\bar{D}$ mesons may become very large at
forward rapidities at RHIC (but not for both simultaneously).

\bef
\psfig{file=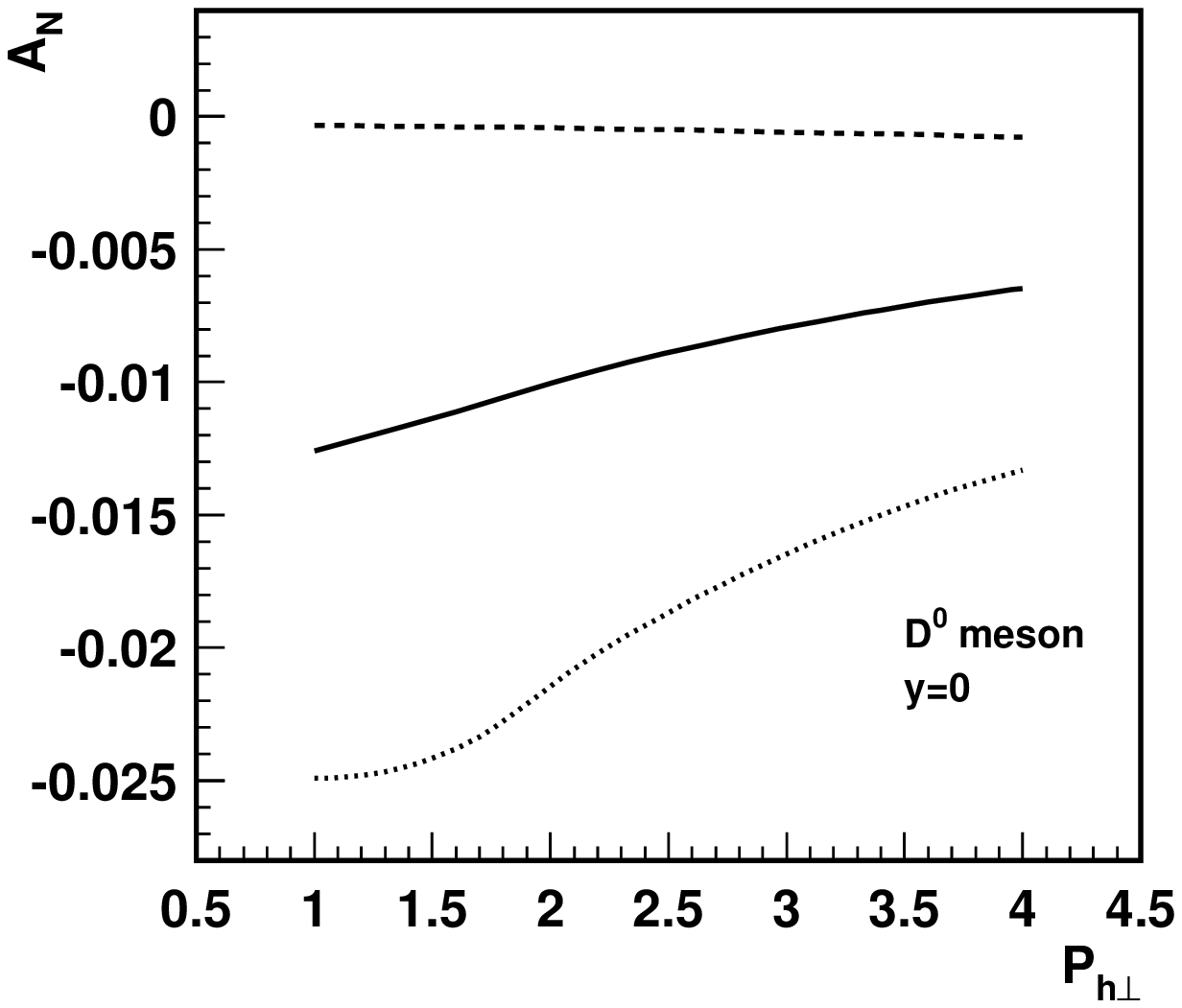,width=2.5in}
\hskip 0.2in
\psfig{file=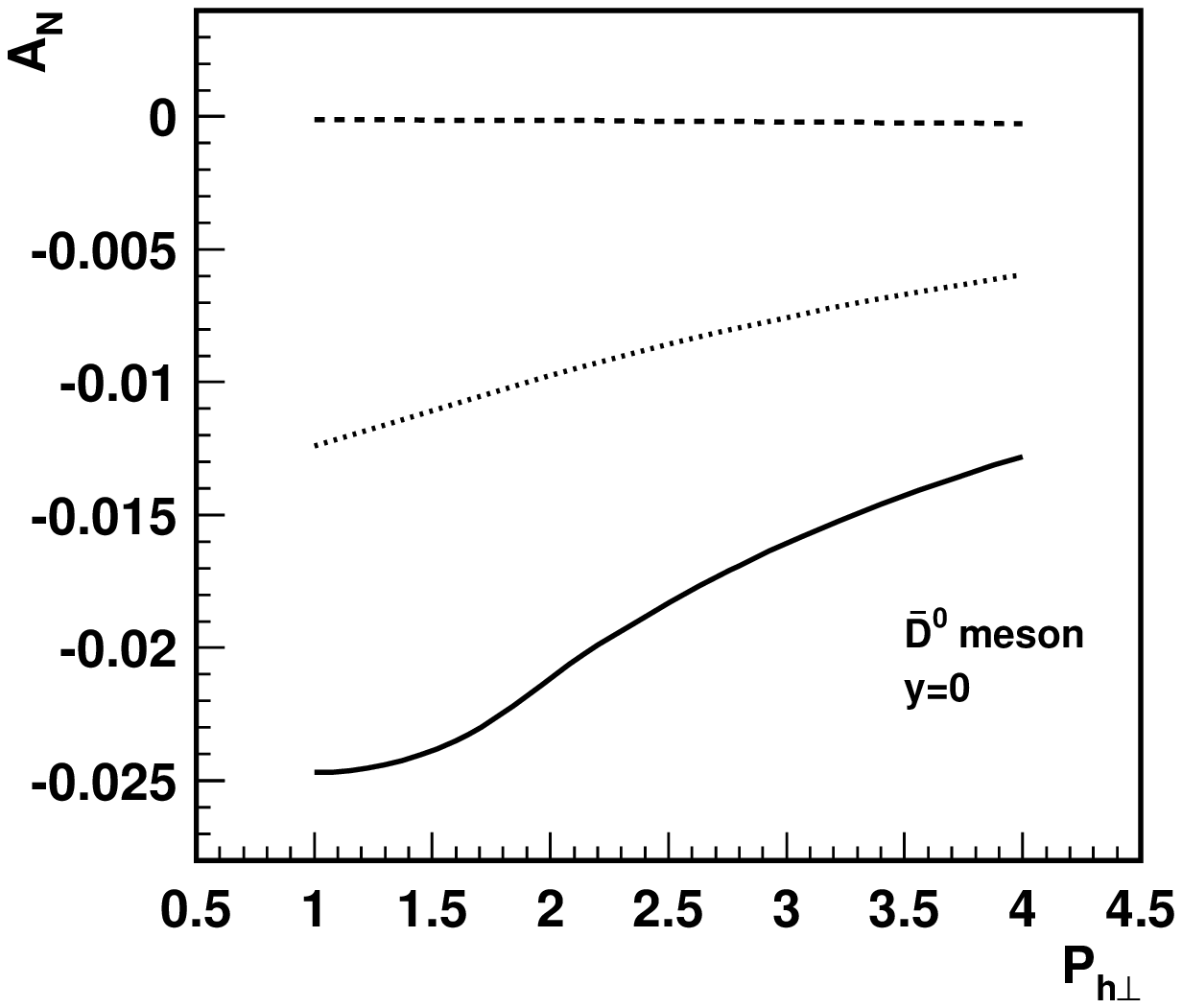,width=2.5in}
\caption{The SSA as a function of $P_{h\perp}$ for $D^0$ (left) and $\bar{D}^0$ mesons (right) at mid-rapidity, $y=0$, 
and $\sqrt{s}=200$ GeV. The curves are: solid ($\lambda_f=\lambda_d=0.07$ GeV), dashed ($\lambda_f=\lambda_d=0$), dotted ($\lambda_f=-\lambda_d=0.07$ GeV).}
\label{pt_dep_mid}
\eef

In Figs.~\ref{pt_dep_mid} and \ref{pt_dep_forward}, we show $A_N$ 
for $D^0$ and $\bar{D}^0$ meson production as a function of $P_{h\perp}$,  
at mid-rapidity ($y=0$) and forward-rapidity ($y=1.8$), respectively. The 
absolute values of the SSAs decrease as a function of $P_{h\perp}$, 
which is a natural behavior of the twist-3 effect in QCD collinear 
factorization.  As before, while the contribution by the 
quark-gluon correlation functions is very small, the two tri-gluon 
correlation functions can make sizable, and very different, contributions 
to the SSA, thanks to the difference in the partonic hard parts in 
Eq.~(\ref{DtoDbar}).

\bef
\psfig{file=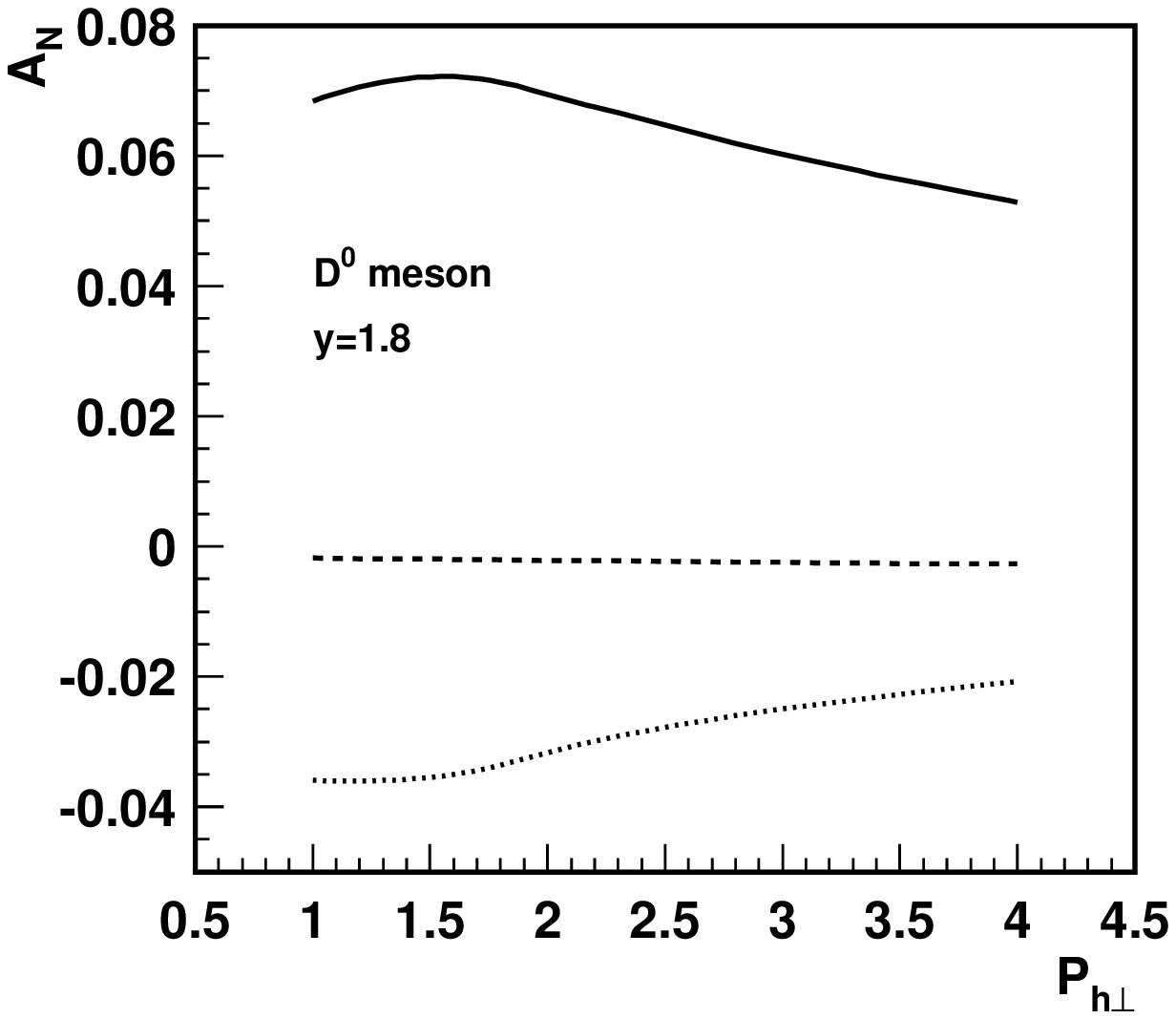,width=2.5in}
\hskip 0.2in
\psfig{file=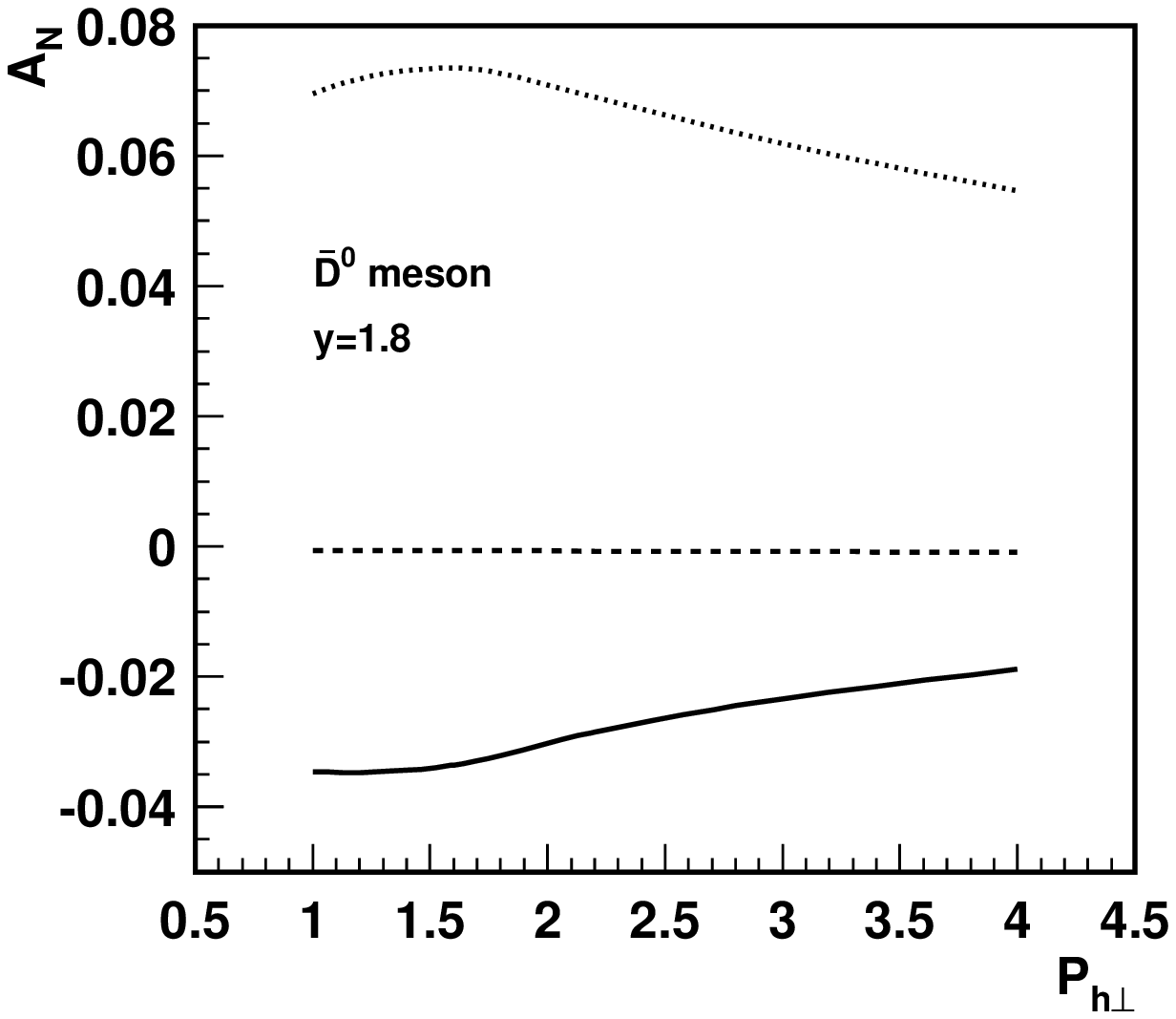,width=2.5in}
\caption{Same as Fig.~\ref{pt_dep_mid}, but at forward rapidity, $y=1.8$.}
\label{pt_dep_forward}
\eef

We finally note that first experimental data on the SSAs for open-charm
production are now emerging from RHIC \cite{liu}. The uncertainties are
currently too large to allow distinction of the various models we
have proposed, but the measurements are certainly very encouraging.

\section{Summary and Conclusions}
\label{ssa_con}

We have studied the single transverse-spin asymmetries for $D$ and $\bar{D}$ 
meson production in hadronic collisions. Within the QCD collinear 
factorization approach, we have calculated both the derivative and the 
non-derivative contributions to the SSAs. We have included the contributions 
by the twist-3 quark-gluon correlation functions, shown in Eq.~(\ref{qqtoD}), 
as well as by the twist-3 tri-gluon correlation functions, given 
in Eq.~(\ref{ggtoD}). We have found that the quark-gluon correlation 
functions alone generate a negligible SSA for open charm production in 
almost all the kinematic region of interest at RHIC.  
We thus conclude that any sizable 
SSAs observed in $D$ or $\bar{D}$ production will be a 
discovery of tri-gluon correlations inside a polarized proton. 

We have proposed several simple models for the two unknown tri-gluon 
correlation functions and used these to make predictions for the SSAs 
in $D$ meson and $\bar{D}$ meson production in polarized proton-proton 
collisions at RHIC. We have found that within our models, the asymmetries 
could be sizable and provide an important source of information on the 
two unknown tri-gluon correlation functions. We also found that the two 
correlation functions play a very different role in generating SSAs for 
$D$ and $\bar{D}$ mesons, due to the sign difference between
the partonic hard parts 
for charm and anti-charm production (see Eq.~(\ref{DtoDbar})).
Comparison of the SSAs for $D$ and $\bar{D}$ mesons could thus 
provide an excellent tool for separating the two tri-gluon correlation 
functions. With data for the SSAs in open charm production arriving 
from RHIC \cite{liu}, we will be able to learn for the first time about
the dynamics of multi-gluon correlations inside a polarized proton.

We close this paper with a theoretical observation. As we 
mentioned in the Introduction, two approaches are commonly used 
for the study of SSAs in high energy collisions: 
the collinear factorization we have used in this paper, and the 
transverse momentum dependent (TMD) factorization approach. The
two are known to be closely connected and complementary to each 
other \cite{UnifySSA}.  As was shown in Ref.~\cite{mulders}, 
the twist-three quark-gluon correlation function, $T_{q, F}(x,x)$, 
is related to a moment in transverse momentum of the corresponding quark 
Sivers function, $q_T(x,k_\perp)$:
\ben
T_{q, F}(x,x)=\frac{1}{M_p}\int d^2\vec{k}_\perp\,
\vec{k}_\perp^2\, q_T(x,k_\perp)\, ,
\label{TF2Sivers}
\een
up to ultraviolet renormalization. Here $M_p$ is a hadronic mass scale.  
Applying the same method used to derive the above relation to the 
TMD gluon distribution, we would obtain an identical relation between the 
tri-gluon correlation function $T_G^{(f)}(x,x)$ and the moment of the 
gluon Sivers function.  The occurrence of the antisymmetric color 
contraction of the three gluonic field strengths in the tri-gluon correlation 
function is a natural consequence of the expansion of the gauge link in 
the adjoint representation when taking the transverse momentum moment of the 
gluon Sivers function. It is interesting to note that the other tri-gluon 
correlation function, $T_G^{(d)}(x,x)$, does not appear to have a simple 
relation connecting it to the operator definition of the gluonic Sivers 
function. That said, in the TMD approach one can define tri-gluon
correlators with both the $f$ and $d$ color structures~\cite{cedran}
when considering general QCD hard-scattering processes, which correspond
to our $T_G^{(d)}$ and $T_G^{(f)}$ after transverse-momentum
weighting. In any case, both $T_G^{(d)}(x,x)$ and $T_G^{(f)}(x,x)$ 
are allowed and in general present in the collinear factorization approach.  
It is therefore very important to measure the difference of SSAs in 
$D$ and $\bar{D}$ meson production, in order to identify the role of 
$T_G^{(d)}(x,x)$.

\section*{Acknowledgments}

We thank G. Kramer for providing us with their Fortran code for the 
$D$ meson fragmentation functions. Z.K. and J.Q. are supported in part 
by the U. S. Department of Energy under Grant No.~DE-FG02-87ER40371. 
J.Q. thanks the Institute of High Energy Physics, Chinese Academy
of Science for its hospitality during the completion of this work.
W.V. and F.Y. are grateful to the U.S. Department of Energy 
(contract number DE-AC02-98CH10886) for providing the facilities essential 
for the completion of this work.  F.Y. is also supported in part by the 
U.S. Department of Energy under contract DE-AC02-05CH11231.


\end{document}